\documentclass[twocolumn,prl,aps,showpacs,superscriptaddress]{revtex4}
\usepackage{epstopdf}
\usepackage{dcolumn}
\usepackage{bm}
\usepackage{epsfig}
\usepackage{pictex}
\begin{document}
\title{Coulomb parameters and photoemission for the molecular metal TTF-TCNQ}
\author{Laura Cano-Cort\'es}
\affiliation{Departamento de F\'isica Te\'orica de la Materia Condensada, 
             Universidad Aut\'onoma de Madrid, 28049 Madrid, Spain}
\author{Andreas Dolfen}
\affiliation{Institut f\"ur Festk\"orperforschung, Forschungzentrum J\"ulich, 
             52425 J\"ulich, Germany}
\author{Jaime~Merino}
\affiliation{Departamento de F\'isica Te\'orica de la Materia Condensada, 
             Universidad Aut\'onoma de Madrid, 28049 Madrid, Spain}
\author{J\"org~Behler}
\affiliation{Fritz-Haber-Institut der Max-Planck-Gesellschaft, Faradayweg 4-6,
             14195 Berlin, Germany}
\author{Bernard~Delley}
\affiliation{Paul-Scherrer-Institut, HGA/123, CH-5232 Villigen PSI, Switzerland}
\author{Karsten Reuter}
\affiliation{Fritz-Haber-Institut der Max-Planck-Gesellschaft, Faradayweg 4-6,
             14195 Berlin, Germany}
\author{Erik Koch}
\email{E.Koch@fz-juelich.de}    
\affiliation{Institut f\"ur Festk\"orperforschung, Forschungzentrum J\"ulich, 
             52425 J\"ulich, Germany}
\date{\today}
\begin{abstract}
We employ density-functional theory to calculate realistic parameters for
an extended Hubbard model of the molecular metal TTF-TCNQ. Considering
both intra- and intermolecular screening in the crystal, we find
significant longer-range Coulomb interactions along the molecular stacks, 
as well as inter-stack coupling. We show that the long-range Coulomb term
of the extended Hubbard model leads to a broadening of the spectral density, 
likely resolving the problems with the interpretation of photoemission
experiments using a simple Hubbard model only. 
\end{abstract}
\maketitle

As the first realization of a molecular metal, the quasi one-dimensional
molecular crystal TTF-TCNQ has been studied thoroughly for more than thirty 
years \cite{Heeger}. The different electro-negativity of the two molecules leads to
charge transfer of about $0.6e$ from TTF to TCNQ, effectively doping 
molecular stacks of like molecules, which become conducting. Low dimensionality and
strong Coulomb repulsion \cite{Kagoshima} lead to many-body effects
which appear in spin susceptibility \cite{Torrance} and angular-resolved
photoemission experiments (ARPES) \cite{Jerome,Ito05}. Most prominently, TTF-TCNQ 
is one of the few systems in which spectroscopic signatures of spin-charge 
separation have been clearly observed \cite{Sing}. The interpretation of the
measured spectra relies, however, on rough estimates of the Coulomb parameters, which
come from early experimental \cite{Torrance} and theoretical \cite{Hubbard,Mazumdar}
works. They suggest that the Coulomb repulsion $U$ between two electrons on a 
molecule is larger than the band-width $W$. Yet, despite their importance for a
correct description of the electronic properties, more quantitative values for the
Coulomb parameters of TTF-TCNQ are to date not available.

In addition, a description of TTF-TCNQ in terms of a pure one-dimensional 
Hubbard model with only on-site Coulomb interaction seems to have problems. 
For the interpretation of 
recent ARPES data \cite{Sing} the hopping matrix element $t$ 
within a molecular chain has to be taken about twice of what 
is estimated from band structure \cite{Fraxedas} and experiment 
\cite{Kagoshima}. Moreover, using these parameters, the temperature
dependence of the spectral function cannot be understood \cite{Assaad,Maekawa06}. 
These calculations, however, neglect the inter-molecular Coulomb 
repulsion $V,$ the relevance of which was already suggested by Hubbard 
\cite{Hubbard} and through diffusive X-ray scattering experiments, 
which display Wigner-type fluctuations \cite{Kagoshima}. 

Our aim in this Letter is to provide a realistic model for the description 
of TTF-TCNQ and to understand the spectral function. To calculate the 
screened Coulomb parameters for an extended 
Hubbard model, we proceed in two steps. Using density-functional theory (DFT),
we first determine the screening due to the electrons on the same molecule 
(intra-molecular screening). To also include the inter-molecular screening we 
describe the other molecules in the lattice by point polarizabilities and 
check this electrostatic approach against constrained-DFT calculations for small 
clusters. Our calculations yield a sizable coupling between neighboring molecules 
on different stacks, as well as important longer-range Coulomb interactions $V$
along the stacks. Focusing on the effect of $V$ on the spectral density, we find 
that it leads to a broadening consistent with photoemission experiments.

\begin{figure}
\begin{center}
\epsfig{file=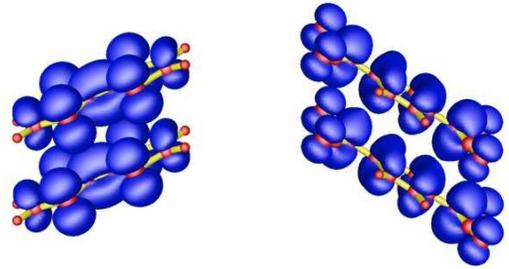,width=7.0cm,clip=}\\
\end{center}
\vspace{-3ex}
\caption{\label{fig1}(Color online) 
TTF-HOMO (left) and TCNQ-LUMO (right) of two neighboring molecules 
in TTF-TCNQ. Since both are planar $\pi$-orbitals, the molecules 
facing each other in a stack are quite close. This gives rise to 
the hopping $t$ and also implies a strong nearest-neighbor Coulomb 
repulsion $V.$}
\end{figure}

\newcommand{\Vbare}{V_\mathrm{bare}}
\newcommand{\Ubare}{U_\mathrm{bare}}

\begin{table*}
\vskip 0.1in
\begin{tabular}{lll@{\hspace{6ex}}lll@{\hspace{6ex}}lll@{\hspace{6ex}}lll}
        &TTF&TCNQ&        &TTF&TCNQ&         &TTF&TCNQ&          &TTF&TCNQ\\
\hline
$\Ubare$&5.9&5.4 &$\Vbare$&3.1&2.9 &$\Vbare'$&1.8&1.7 &$\Vbare''$&1.2&1.2\\
$U_0$   &4.7&4.3 &$V_0$   &2.9&2.8 &$V_0'$   &1.8&1.7 &$V_0''$   &1.2&1.2\\
$U$     &2  &1.7 &$V$     &1  &0.9 &$V'$     &0.55&0.4&$V''$     &0.4&0.3\\
\end{tabular}
\caption{\label{tableI}
Hubbard parameters for TTF-TCNQ. 
$\Ubare$ is the direct Coulomb integral, $U_0$ includes the renormalization
due to intra-molecular screening, and $U$ is the screened on-site Coulomb
term in the crystal. Correspondingly, $V$ gives the first, $V'$ the second, 
and $V''$ the third nearest-neighbor Coulomb interaction between molecules 
on the same TTF or TCNQ stack. All energies are in eV. For comparison, the 
band-width is about 0.7~eV.}
\end{table*}

We start our calculation of the Coulomb matrix elements from the bare 
Coulomb integrals between molecular orbitals (MOs) on two molecules a
distance $l$ apart
$ \Vbare^l=\int\!d^3 {\bf r} \int\!d^3 {\bf r'} 
  \rho_0({\bf r }) \rho_l({\bf r'})/|{\bf r}-{\bf r'}|$, 
with $\Vbare^0=\Ubare$. We calculate the charge density $\rho({\bf r})$ of 
the highest occupied molecular orbital (HOMO) of TTF and the lowest 
unoccupied MO (LUMO) of TCNQ from all-electron density-functional theory
(DFT) using Gaussian orbitals \cite{NRLMOL} and the Perdew-Burke-Ernzerhof
functional \cite{PBE}. The results for the on-site, $\Ubare$, and the first, 
$\Vbare$, second, $\Vbare'$, and third, $\Vbare''$, nearest neighbor 
Coulomb terms on the TTF/TCNQ stacks are listed in Table \ref{tableI}. 
We notice that $\Vbare$ is substantial, being of the order
of $\Ubare/2$. This can be understood from Fig.~\ref{fig1}: 
The $\pi$-molecular orbitals stick out of the molecular
planes so that the charges from the two molecules repel strongly.
For distances $l$ larger than twice the molecular radius, 
$\Vbare^l$ is given by $e^2/l$.
As a side-product we obtain from the bonding-antibonding
splitting for a pair of neighboring molecules the hopping matrix elements
along the stacks: $|t|=0.15$~eV for TTF and 0.18~eV for TCNQ.

The on-site Coulomb parameter $U_0$ including intra-molecular screening is 
obtained by calculating the DFT total energy of an isolated molecule, putting
charge $q$ in the HOMO or LUMO, respectively, and fitting to 
$E_U(q)=a_0+a_1 q+U_0 q^2/2$. The results are listed in Table~\ref{tableI}
and agree well with previous calculations for TCNQ \cite{Johansen}. 
We note that intra-molecular screening reduces $\Ubare$ by more than 1~eV.
In order to obtain the Coulomb parameters $V_0^l$, we calculate the total energy
of two molecules at distance $l$, putting charge $q/2$ on each and fitting 
to $E_V(q)=2E_U(q/2)+b_0+b_1 q+V_0 q^2/4$. In contrast to the on-site term, 
we find hardly any screening of the $\Vbare$'s, since the molecules mainly screen 
their own charge.

Determination of the inter-molecular screening contribution requires 
calculations of the energy of an infinite lattice of molecules. Constraining
a charge $q$ to a single molecule allows to obtain $U$, while constraining it
to two molecules at a distance $l$ provides the different $V_l.$ Unfortunately,
such constrained DFT calculations \cite{constrDFT} can only be done for very
small clusters of molecules. In order to reach the infinite-system limit, 
we therefore evaluate the reliability of approaches which represent the molecules 
by their polarizabilities. To determine the polarizability tensor of the isolated
molecules, we calculate, with DFT, the dipole moments in a homogeneous external
field along the principal axes and extract the linear response. The results are
given in Table \ref{alphas} and are consistent with estimates derived from the
experimental dielectric constant through the Clausius-Mossotti relation.
\begin{table}
\begin{tabular}{l@{\hspace{3ex}}r@{\hspace{3ex}}r@{\hspace{3ex}}r}
      & $\alpha_{xx}$ & $\alpha_{yy}$ & $\alpha_{zz}$\\ \hline
 TTF  & 226           & 160           & 88\\
 TCNQ & 440           & 184           & 82\\
\end{tabular}
\caption{\label{alphas}
Principal elements of the polarizability tensor of the TTF and the TCNQ 
molecule in atomic units (bohr$^3$). $x$ ($y$) is the long (short) axis of the
molecule, while $z$ is perpendicular to the molecular plane.}
\end{table}

One might then replace a molecule in the lattice by a point polarizability 
at the molecule's center of gravity. Such an approach, which works nicely,
e.g., in the case of C$_{60}$ \cite{C60}, fails in the present case, since
the molecules in a TTF/TCNQ stack come so close that they are inside the
convergence radius of the dipole approximation. To avoid this problem we 
distribute the polarizability uniformly over the non-hydrogen atoms of the 
molecule \cite{mazur}.
To test the reliability of this approach, we compare the 
screening due to the nearest molecules, where the distributed dipole approach
should work least well, to constrained DFT calculations, using an implementation 
of the method in DMol$^3$ \cite{Behler}. 
Calculating the reduction $\delta U$ of the on-site Coulomb term due to 
screening of just the two nearest neighbors of a TTF molecule in a 
stack, we find $\delta U=1$~eV, while the distributed-dipole approach 
yields 0.9~eV. A similarly good agreement is found when taking also the 
second nearest neighbors on the stack into account: The constrained DFT 
calculation for this 5 molecule system gives $\delta U=1.7$~eV, compared 
to 1.6~eV from the distributed-dipole approach. This points at a quite small
error, in particular when considering that the screening due to the more distant
molecules along the stack will be increasingly well described by the
dipole approximation. 

We therefore employ the distributed-dipole approach to compute the screening
of the Coulomb interaction in a lattice of polarizable points with the
experimental crystal structure of Ref.~\cite{lattice}. We use the notation 
of Allen \cite{Allen04}, where a configuration of dipole moments on the lattice 
is denoted by a vector $|{\bf p}\rangle$ and the dipole-dipole interaction on 
the lattice is described by a matrix $\bf\Gamma$. If the dipoles arise via
polarization, their energy in an external field 
$|{\bf E}^\mathrm{ext}\rangle$ 
is given by 
$\langle{\bf p}|\alpha^{-1}-{\bf\Gamma}|{\bf p}\rangle/2
-\langle{\bf p}|{\bf E}^\mathrm{ext}\rangle$\,,
where $\alpha$ is the polarizability tensor. The configuration that minimizes 
the energy follows from the variational principle: 
$|{\bf p}\rangle=(\alpha^{-1}-{\bf\Gamma})^{-1}|{\bf E}^\mathrm{ext}\rangle$.
As the energy is quadratic in the external field we can focus on pairs of
electrons. Writing the field of two point charges located at lattice sites
$n$ and $m$ as ${\bf E}^\mathrm{ext}={\bf E}_n+{\bf E}_m$, we find that 
screening lowers the energy by
\begin{displaymath}
 \delta W=
  -\langle{\bf E}_n|(\alpha^{-1}-{\bf\Gamma})^{-1}|{\bf E}_n\rangle
  -\langle{\bf E}_m|(\alpha^{-1}-{\bf\Gamma})^{-1}|{\bf E}_n\rangle\,,
\end{displaymath}
where in the first term, which gives the screening of each individual charge,
we have used translational invariance, and in the second term, which describes
the screening of the interaction, we have used inversion symmetry. If the two
point-charges are located at the same position, we get the screening of the 
on-site interaction
$\delta U = -\langle{\bf E}_0|(\alpha^{-1}-{\bf\Gamma})^{-1}|{\bf E}_0\rangle$,
while for $|n-m|=l$ we find 
$\delta V^l=-\langle{\bf E}_0|(\alpha^{-1}-{\bf\Gamma})^{-1}|{\bf E}_l\rangle$.
We have performed these calculations for clusters of up to 400 molecules. 
Extrapolating to the infinite lattice limit gives the results compiled in
Table~\ref{tableI}. We find that for TTF-TCNQ the screening is very
efficient, reducing $U=U_0-\delta U$ by more than 2~eV, while for the longer
range interactions the screening is somewhat smaller.

Comparing to the band-width $W$ of about 0.7~eV \cite{Fraxedas}, we find
that $U/W$ is larger than 2, and also the longer range interaction is
substantial, with the nearest neighbor interaction $V$ being about $U/2$.
Even the interaction between third nearest neighbors is still comparable
to $W$. The parameters for the TTF and TCNQ stacks are quite similar, with
TTF, having smaller hopping and larger Coulomb parameters, being slightly 
more correlated.
While the electronic structure in the absence of inter-stack Coulomb
interaction is essentially one-dimensional, we find that there is also a sizable
coupling between neighboring molecules on different stacks. For two neighboring
TTF stacks we find $\Vbare=1.6$~eV, the same for neighboring TCNQ stacks, 
while the interaction between TTF and TCNQ stacks is about 2.1~eV. Screening 
reduces all these values to about 0.4~eV, which is still comparable to the band-width.

Our results therefore show that a realistic description of TTF-TCNQ requires 
an extended Hubbard model, including both inter-stack, as well as longer-range 
Coulomb interactions along the stacks. This is in stark contrast to existing pure 
one-dimensional Hubbard models with on-site term only ($t$-$U$-model) 
\cite{Sing,Assaad,Maekawa06}. Due to the inter-chain coupling we expect Coulomb-drag 
effects \cite{drag}, giving rise to interesting correlated dynamics of the 
individual stacks. Treating this coupling at mean-field level, one arrives at a 
one-dimensional Hubbard model, with Coulomb-parameters renormalized by the 
metallic neighboring chains. This will somewhat reduce $U$ and $V$, while the 
longer-range interactions are suppressed more efficiently \cite{Hubbard}. 
In the following we study the importance of the longer-range Coulomb interaction 
along the stacks by considering the extended Hubbard model of a single TTF or TCNQ chain
\begin{displaymath}
H=-t\!\!\sum_{\langle ij\rangle,\sigma}
 (c^{\dagger}_{i \sigma} c_{j \sigma} + H.c.)
+ U\!\sum_i n_{i\uparrow} n_{i\downarrow} 
+ V\!\sum_{\langle ij\rangle} n_i n_j
\end{displaymath}
using the values for $U$ and $V$ given in Table \ref{tableI} ($t$-$U$-$V$-model). 
With filling $n\approx$~0.6 and 1.4 and very similar parameters, 
the models for a chain of TCNQ and TTF are closely related 
by an electron-hole transformation, with TTF being somewhat more 
correlated than TCNQ. The following discussion of the spectral function for
a chain of TCNQ and applies therefore correspondingly to TTF.

Since $U$ is more than twice the band-width $W$, it seems safe to assume 
that in a TCNQ-chain double occupancies are completely suppressed. The effect 
of $V$ is then to maximize the distance between occupied sites. For filling 
0.6 this results in a Hubbard-Wigner state with occupations 
\ldots1101011010\ldots with a periodicity of 5 lattice sites \cite{Hubbard}. 
However, such a state would be an insulator with a gap opening with $V$.
This is, indeed, what we find solving the $t$-$U$-$V$ Hubbard chain for 
$U\to\infty$. For $U=1.7$~eV, however, the chain is metallic,
indicating that double occupancies can not be neglected. Indeed, for
$V=0$ we find that the probability $d$ of a doubly occupied site is 
$d\approx0.01$, to be compared to $d=0.09$ for $U=0$. The effect of $V$ is to
further increase $d$. For $V=0.9$~eV we have $d\approx0.027$. To obtain the
same value in the $t$-$U$ model, we would have to double $t$ from 0.18~eV 
to about 0.37~eV. Evidently, the nearest-neighbor Coulomb term encourages
hopping. To understand this, consider what happens when two electrons pass 
each other: first they must become neighbors (energy $V$), then, to pass, they
must form a doubly occupied site (energy $U$), to finally end up in a
configuration where the neighbors have exchanged roles. This process thus
involves effectively an energy $U-V$. This can be translated to an effective
hopping $t_\mathrm{eff}\approx t\,U/(U-V)$ for a corresponding 
$t_\mathrm{eff}$-$U$ model. For our parameters we find 
$t_\mathrm{eff}\approx0.38$~eV.

\begin{figure}
 \centerline{\resizebox{3.2in}{!}{
   \rotatebox{270}{\includegraphics{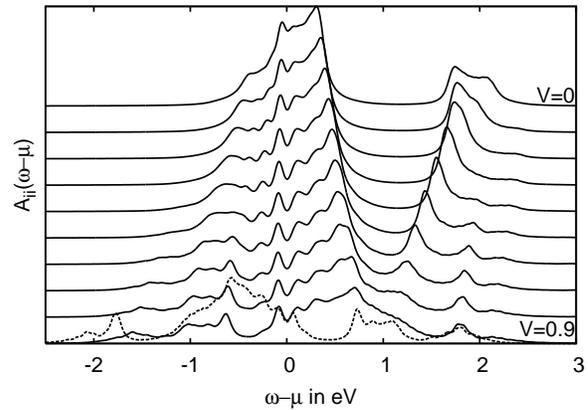}}}}
 \caption{\label{spect}
  Spectral function for the one-dimensional $t$-$U$-$V$ model
  on 20 sites. The full lines show the effect of $V$ for a TCNQ chain with 
  $n=0.6$, $t=0.18$, $U=1.7$, and $V$ increasing from 0 to 0.9~eV. 
  The dotted line is for a TTF chain with $n=1.4$, $t=-0.15$, $U=2.0$, 
  and $V=1.0$~eV.}
\end{figure}
Figure \ref{spect} shows the spectral function $A_{ii}(\omega)$ of the 
$t$-$U$-$V$ chain as function of $V$. We see that the main effect of $V$ is to 
broaden the spectrum around the Fermi level. A similar broadening has also been 
observed in Ref.~\cite{Maekawa06} and is reminiscent of what happens when we 
increase $t$ in the $t$-$U$ model as proposed in Ref.~\cite{Sing}.
Remarkably, the broadening can already be understood in first-order
Rayleigh-Schr\"odinger perturbation theory: If we consider the $V$-term in the 
$t$-$U$-$V$ Hamiltonian as a perturbation, then, to first order, the energy of 
the many-body states changes by $V\sum_{\langle ij\rangle}\langle n_i n_j\rangle$,
while the wave-functions remain unchanged. From the Lehmann representation we 
thus see that the position of the spectral peaks shift linearly in $V$. 
This shift is proportional to $\sum_{\langle ij\rangle}\langle n_i n_j\rangle$, 
which, for the low energy states, tends to increase with the energy.
Thus the further a pole is from the Fermi level, the more it moves with $V$, 
leading to the observed broadening of the spectrum. 

We finally illustrate an interesting effect of the polarizability $\alpha$ on
the screening of the Coulomb interaction. For this, we consider a one-dimensional
chain of polarizable points with lattice spacing $b$, which we can solve 
analytically. For $\alpha<b^3/4\zeta(3)$, where $\zeta(n)$ is the Riemann zeta 
function, such a system is stable against spontaneous polarization, so we can use  
translational invariance of the dipole-dipole interaction matrix to diagonalize 
$1/\alpha-{\bf\Gamma}$ in a plane-wave basis \cite{Allen04}. From the spectral 
representation we then obtain
\begin{displaymath}
 \delta V^l = -\frac{e^2}{2\pi b}\int_{-\pi}^\pi\!\!dk\;cos(lk)\;
  \frac{\Im({\rm Li}_2(e^{ik}))^2}{b^3/4\alpha - \Re({\rm Li}_3(e^{ik}))}\,,
\end{displaymath}
where ${\rm Li}_2(z)$ and ${\rm Li}_3(z)$ is the di- and trilogarithm, 
respectively \cite{polylog}. From this solution we find that the screening is 
less efficient for larger $l$, even showing {\em antiscreening} for $l\ge2$ 
and $\alpha$ not too close to the ferroelectric instability. Approaching
the critical $\alpha$, we find that the interactions $V^l$ become almost
independent of the distance $l$. Thus close to the ferroelectric instability 
electronic correlations are lost.
TTF-TCNQ should be very close to the transition.
Hence it would be exciting to see what happens to the electronic correlations 
in TTF-TCNQ under sufficient hydrostatic pressure.

In conclusion, we have calculated a realistic set of Coulomb parameters for 
TTF-TCNQ. We find that the commonly used values for the on-site interaction $U$ 
are correct, but  that interactions between electrons on neighboring sites are 
significant  and must be taken into account. Including the nearest neighbor 
interaction $V$ on a TCNQ or TTF chain, we find that the spectra are broadened, 
an effect that can also be mimicked by increasing the hopping term $t$. Not having 
to assume an enhanced $t$ to obtain the proper width of the low-temperature spectral 
function, should also lower the estimate of the temperature scale $T_J$, above 
which the signs of spin-charge separation are lost, to the experimentally observed 
range. In addition we have found that the Coulomb terms couple different chains, 
possibly giving rise to interesting Coulomb-drag effects.

\acknowledgments
We thank F.F.~Assaad, V.~Blum, R.~Claessen, F.~Flores, O.~Gunnarsson, M.~Sing, and
H.~V\'azquez for discussions.
J.M.\ is grateful to the Ram\'on y Cajal program of Ministerio de Ciencia y
Tecnolog\'ia, Spain for financial support through contract CTQ2005-09385. 
The Lanczos runs were performed on the J\"ulich BlueGene/L under grant JIFF22.

\end{document}